\documentclass[showpacs,preprintnumbers,amsmath,amssymb]{revtex4}
\usepackage{graphicx}
\usepackage{dcolumn}
\usepackage{bm}
\begin{document}
\title{On the BRST and finite field dependent BRST of a model where vector and
axial vector interaction get mixed up with different weight}

\author{Safia Yasmin}
\affiliation{Indas Mahavidyalaya, Bankura - 722205, West Bengal,
India}
\author{Anisur Rahaman}
\email{1. anisur.rahman@saha.ac.in, 2. manisurn@gmail.com}
\affiliation{Hooghly Mohsin College, Chinsurah, Hooghly - 712101,
West Bengal, India}

\date{\today}

\begin{abstract}
The generalized version of a lower dimensional model where  vector
and axial vector interaction get mixed up with different weight is
considered. The bosonized version of which does not posses the
local gauge symmetry. An attempt has been made here to construct
the BRST invariant reformulation of this model using Batalin
Fradlin and Vilkovisky formalism. It is found that the extra field
needed to make it gauge invariant turns into Wess-Zumino scalar
with appropriate choice of gauge fixing. An application of finite
field dependent BRST and anti-BRST transformation is also made
here in order to show the transmutation between the BRST symmetric
and the usual non-symmetric version of the model.
\end{abstract}
 \maketitle

\section{{\bf Introduction}}
Dynamical equations of physical system cannot always be described
in terms of  observable physical degrees of freedom which pose
problem to the straightforward physical interpretation of  the
solution of  evaluation equations \cite{DIR, SU, DMG}.
   In some cases, few solutions need to be excluded since they do not describe the real physical situation or it may be the case that certain class of apparently different solutions appears to be physically indistinguishable. The BRST-formalism \cite{BRS1, BRS2, BRS3} has been developed precisely to deal with such systems.  It is a technique to enlarge the phase space of a gauge theory and to restore the symmetry of the gauge fixed action in the extended phase space keeping the physical contents of the theory intact. To study the unitarity and renormalization it is instrumental. The unphysical ghost field acquires prominent status rendering its valuable service in bringing back the symmetry of the gauge fixed effective action with the preservation of the fundamental unitary property. Since this symmetry mixes all the fields (physical and ghost) in such a way that ghost field along with the other fields need to be treated on the same footing and that  forces to regard the ghost field along with all the other field as a different component of a  single geometrical object.

The role field dependent BRST (FFBRST) is almost similar to the
BRST so far symmetry is concerned. It does protect nilpotency and
reflects the symmetry of the gauge fixed action \cite{JOG} of a
physically sensible theory. It can be considered as a
generalization over the usual BRST formalism where transformation
parameters becomes finite, field dependant and anti-commuting in
nature \cite{JOG}. Unlike BRST transformation, it fails to keep
the measure of the generating functional unchanged \cite{JOG}.
However, the change appeared there renders several important
services to make an equivalence between the different effective
actions of a particular theory \cite{JOG}.
 In this context, the services  obtained through the exploitation of the change entered into  the measure of the generating functional to relate the different gauge fixed actions of a particular theory is remarkable \cite{JOG}. BRST and FFBRST are therefore equally important and interesting in their own right. So application of BRST as well as FFBRST formalism on any physically sensible theories would be of considerable interest and would certainly add a new contribution to the formal field theoretical regime.

In this context, we consider a (1+1) dimensional generalized
version of Schwinger (GVSM) model where axial and vector
interaction get mixed up with different weight \cite{BAS1,
BAS2,BAS3, ARSN}. The most interning feature of the model is its
ability to interpolate the two most important lower dimensional
field theoretical models through its mixing weight factor: the
models are well celebrated vector Schwinger model \cite{SCH} and
its chiral generation, commonly known as chiral Schwinger mode
\cite{JR}. Schwinger model started a glorious journey for its
potential of describing the mass generation along with its ability
to describe the confinement aspect of fermion in lower dimension
and has been extensively studied over the years \cite{LO,
MIA1,MIA2,PSP1,PSP2,RA,AR1,AR2}. Chiral generalization of this
model too has been studied with great interest after the removal
of its unitarity problem by Jackiw and Rajaraman
\cite{JR,GIR1,GIR2,KH,
PM,PMS,SM,SM1,AR2,AR3,AR4,AR5,AR6,AR7,AR8,AR9,APR1,APR2,ARANN1,ARANN2}.

Recently, an attempts has been made by us to quantize both the
gauge invariant and gauge non-invariant version of GVSM
\cite{ARSN}. This model in its bosonized version does not possess
the local gauge symmetry, since it becomes essential to take into
account the anomaly to protect the unitarity of this model. Here
mass generation takes place indeed, via a kind of dynamical
symmetry breaking. However, unlike Schwinger model \cite{SCH},
here the fermions are found to get liberated which may be
considered as de-confinement phase of fermions. We should mention
here that the fermion are found to remain confined when the model
turns into Schwinger model in absence of its axial interaction
part. The model provides so many interesting and surprising
insights into the phenomena related to mass generation and
confinement aspect of fermion, charge shielding etc., that till
now it remains as a subject of several interest.  So naturally,
the extension of the model coined in \cite{BAS1}, which has the
ability of combining these two model s into a single structure
would be of worth investigations. Besides, in order to protect
unitarity, inclusion of anomaly becomes essential and it adds
further interest in another direction, because one loop correction
enters there holding the hand of anomaly. But it certainly breaks
the local gauge symmetry. So the study related to the restoration
of symmetry would be instructive which we have attempted in the
present work. We, therefore, in our present work consider the GVSM
and attempts is made here towards the reformulation of  a BRST
invariant effective action by the use of Batalin, Fradkin and
Vilkovisky (BFV) formalism. The scheme developed by Batalin,
Fradkin and Vilkovisky towards the conversion of a set second
class constraint into first class set helps to get this symmetric
transmuted form. It is known that for the above transmutation some
extra fields are needed. These fields are known as auxiliary
fields. These auxiliary fields turn into Wess-Zumino scalar with
appropriate choice of gauge fixing conditions for some favorable
situations. So at first an attempt has been made here towards the
BRST invariant reformulation  of this model using BFV formalism
\cite{FV,BF,FIK,BT,IA}.  In fact, we have used the improved
version presented by Fujiwara and Igarishi and Kubo (FIK)
\cite{FIK}, since it is known that it generally helps to obtain
the Wess-Zumino \cite{WESJ} action associated with the model in
most of the cases \cite{YW, YWS, MI, SG, ARS}.

Application of FFBRST formalism on this model would also be
instructive like its ancestor BRST formalism and would add a new
contribution to formal field theoretical regime. So an extension
using FFBRST formulation is also made here to show how the
contribution that enterers into the  measure of the generating
functional under FFBRST  transformation helps to  convert  the
BRST invariant effective action into its original gauge
non-invariant version to ensures that the physical contents of
these two effective actions are identical.  The recent works
\cite{BHS, BHS1, BHS2, BHS3, BHS4, BHS5, BHS6, BHS7, BHS8, BHS9,
BHS10}, indeed provides much insight intothe way of approach
towards  our recent attempt. It reminds the work of Falck and
Kramer \cite{FALCK}, where they explicitly showed that physical
content of chiral Schwinger model \cite{JR} remains identical
both in the usual gauge non invariant action and the gauge
symmetric action of the extended phase space. But it has to be
kept in mind that in that situation the symmetry that was handled
was the local gauge symmetry.

   The paper is organized in the following manner. In Sec. II, we
   have given a brief introduction of the model. Sec. III is devoted to the
   BRST invariant reformulation of the model. In Sec. IV,  FFBRST and
   anti-FFBRST formulation is applied to this BRST invariant effective
   action to get back the original gauge non-invariant form of the
   action through the incredible service of the field dependent
    parameter of FFBRST and anti-FFBRST

\section{{\bf Brief review of the model}}
 The model where we find the mixing of both vector and axial vector interaction with different
 weight is given by the following generating functional
\begin{equation}
Z(A)=\int{d\psi} {d\bar{\psi}} exp[i \int{d^2x}{\cal L}_F],
\label{INT}
\end{equation}
with
 ${\cal
L}_{F}=\bar{\psi}\gamma^{\mu}[i\partial_{\mu}+e\sqrt{\pi}A_{\mu}(1-r\gamma_{5})]\psi.
$ The integration over the fermionic degrees of freedom leads to a
determinant and if  that fermionic determinant is expressed in
terms of auxiliary scalar field $\phi$,  we get
\begin{equation}
Z(A)=\int{d\phi} exp[i \int{d^2x}{\cal L}_B], \label{INT1}
\end{equation}
where  ${\cal L}_{B}=\frac{1}{2}
\partial_{\mu}\phi\partial^{\mu}\phi+eA^{\mu}(\tilde{\partial}_{\mu}+r\partial_{\mu})\phi+\frac{1}{2}a
e^{2}A_{\mu}A^{\mu}.$ Here $a$ is the regularization ambiguity
emerged out during the process of regularization to remove the
divergence of the fermionic determinant. If we now introduce the
kinetic term of the back ground electromagnetic field we will get
the total lagrange density:
\begin{equation}
{\cal
L}_{B}=\frac{1}{2}\partial_{\mu}\phi\partial^{\mu}\phi+eA^{\mu}(\epsilon_{\mu\nu}\partial^{\nu}
+rg_{\mu\nu}\partial^{\nu})\phi+\frac{1}{2}a
e^{2}A_{\mu}A^{\mu}-\frac{1}{4}F_{\mu\nu}F^{\mu\nu}. \label{INT2}
\end{equation}
The Euler-Lagrange equations for the fields describing the
lagrangian density (\ref{INT2}) are
\begin{equation}
\partial^{\mu}F_{\mu\nu}=-a e^{2}A_{\nu}-e(\epsilon_{\mu\nu}\partial^{\mu}\phi
+r g_{\mu\nu}\partial^{\mu}\phi),\label{INT3}
\end{equation}
\begin{equation}
\Box{\phi}=-e(r
g_{\mu\nu}\partial^{\nu}+\epsilon_{\mu\nu}\partial^{\nu})A^{\mu}.
\label{INT4}
\end{equation}
It is known that the most general solution for $A_{\mu}$ is
\begin{equation}
A_{\mu}=\frac{1}{ae^{2}}[r\partial_{\mu}\phi+(a-r^{2})\tilde{\partial_{\mu}}\phi
+(1+a-r^{2})\tilde{\partial_{\mu}}h], \label{INT5}
\end{equation}
and the theoretical spectra are given by
\begin{equation}
(\Box+m^{2})\sigma=0, \label{INT6}
\end{equation}
\begin{equation}
\Box{h}=0, \label{INT7}
\end{equation}
where
\begin{equation}
\sigma=\phi+h,
\end{equation}
and $m^{2}$ is given by
\begin{equation}
m^{2}=\frac{e^{2}a(1+a-r^{2})}{(a-r^{2})}. \label{INT8}
\end{equation}
So the physical subspace of the model is constituted with a
massive boson with square of the mass
$m^{2}=\frac{e^{2}a(1+a-r^{2})}{(a-r^{2})}$ and a massless boson.
In short, this the physical content of the model.

\section {BRST invariant reformulation of GVSM using BFV formalism}
To make the paper self contained let us start with the brief
introduction of the BFV formalism. Consider a phase space of
canonical variables $ q^{i},p_{i} (i=1,2................n)$ in
terms of which the canonical Hamiltonian is $ H_{c}(q^{i},p_{i})$
and the constraints $w_{i}(q^{i},p_{i})$ are embedded there in.
The algebra between the constraints themselves and with the
canonical hamiltonian  respectively are
\begin{equation}
[w_{a},w_{b}]=iw_{c}U^{c}_{ab},  [H_{c},w_{a}]=iw_{b}V^{a}_{b},
 \end{equation}
where  $U^{c}_{ab}$ and $V^{a}_{b}$ are the structure
coefficients. To extract out the physical degrees of freedom, $N$
number of additional conditions $\phi^{a}=0$, are needed  to be
imposed. The constraints $\phi^{a}=0$ and $w_{a}=0$ together with
the Hamiltonian equations may be obtained from the action
\begin{equation}
S=\int{[p_{i}\dot{q^{i}}-H(p_{i},q^{i})-\lambda^{a}w_{a}+\pi_{a}\phi^{a}]}dt,
\end{equation}
where $\lambda^{a}$  and $\pi_{a}$ are lagrange multiplier having
Poisson's bracket
 $[\lambda^{a},\pi^{a}]=i\delta^{a}_{b}$ and the gauge fixing
 conditions  contain $\lambda^{a}$ in the form $\phi^a=\lambda^{a}+ \chi^a
 $.
In order to make the equivalence with the initial theory, we may
introduce  two sets of canonically conjugate anti-commuting ghost
coordinates and momenta $ (C^{i} \bar{P_{i}})$ and $(P^{i}
,\bar{C_{i}})$ having the algebra
$[C_{i},\bar{P_{i}}]=i\delta(x-y)$and $[P^{i},
\bar{C_{i}}]=i\delta(x-y)$. The quantum theory, therefore, can be
given by the generating functional
\begin{equation}
Z_G=\int dq^i dp_1 d\lambda^a d\pi_a dC^a d\bar{P}_a dP^a
d\bar{C}_a e^{iS_G},
\end{equation}
where the action $S_G$ is
\begin{equation}
S_G=\int{[p_{i}\dot{q^{i}}+\bar{P^{i}}C_{i}+\bar{C^{a}}\dot{P_{a}}-H_{m}+\dot{\lambda^{a}}\pi_{a}
+i[Q,G]}dt.
\end{equation} \label{BM}
Here $H_{m}$ is usually known as the minimal hamiltonian, Q is the
BRST charge and G is the gauge fixing function and these are
defined as follows
\begin{equation}
H_m= H_c + \bar{P}_aV^a_bC^b, \label{IM}
\end{equation}
\begin{equation}
Q=C^a\omega_a-\frac{1}{2}C^bC^cU^a_{cb}\bar{p}_a + P^a\pi_a,
\end{equation}
\begin{equation}
G=\bar{C}_a\chi^a + \bar{P}_a\chi^a.
\end{equation}
Let us now proceed to apply the above formalism to the model
considered here for BRST invariant reformulation. The bosonized
lagrangian density for this theory is
\begin{eqnarray}
{\cal L}_{B}&=& \frac{1}{2}\dot{(\phi^2}-\phi^{\prime^{2}})+e(A_{0}{\phi^{\prime}}-A_{1}{\dot\phi)}+er(A_{0}\dot{\phi}-A_{1}\phi^{\prime})\nonumber\\
&+& \frac{e^{2}}{2}a(A_{0}^{2}-A_{1}^{2})
+\frac{1}{2}(\dot{A_{1}}-A_{0}^{\prime})^{2}.\label{LAG}
\end{eqnarray}
 We are now in a state  to proceed towards the BRST invariant reformulation of the
 lagrangian given in (\ref{LAG}).
 In order to proceed to that end, we need to know the constraint structure of the
  theory described
 by the lagrangian (\ref{LAG}).
The momentum corresponding to the fields  $ \phi,A_{0}$ and $A_{1}
$  respectively are
\begin{equation}
\pi_{\phi}=\dot{\phi}-eA_{1}+erA_{0},\label{LAG1}
\end{equation}
\begin{equation}
\pi_{0}=0,\label{LAG2}
\end{equation}
\begin{equation}
\pi_{1}=\dot{A_{1}}-A_{0}^{\prime}. \label{LAG3}
\end{equation}
The equation $\omega_{1}=\pi_{0}=0$, is identified as the primary
constraint of the theory. By the Legendre transformation we obtain
the following canonical Hamiltonian:
\begin{eqnarray}
H_{c} &=& \int dx[\pi_{\phi}\dot{\phi}+\pi_{1}\dot{A_{1}}
-(\frac{1}{2}\dot{(\phi^2} -\phi^{\prime^2})+e(A_{0}\phi^{\prime}
-A_{1}{\dot\phi)}\nonumber\\
&+&
er(A_{0}\dot{\phi}-A_{1}\phi^{\prime})+\frac{e^{2}}{2}a(A_{0}^{2}
-A_{1}^{2})+\frac{1}{2}(\dot A_{1} -A_0')^2)].\label{LAG4}
\end{eqnarray}
After a little algebra we find that (\ref{LAG4}) reduces to
\begin{eqnarray}
H_{c}&=&\frac{1}{2}(\pi_{\phi}^{2}+\pi_{1}^{2}+ \phi'^2)
+\pi_{1}A_{0}^{\prime}
+e\pi_{\phi}(A_{1}-rA_{0})\nonumber\\
&-&\frac{e^{2}}{2}a(A_{0}^{2}
-A_{1}^{2})+\frac{1}{2}e^{2}(A_{1}-rA_{0})^{2}+
e\phi^{\prime}(rA_{1}-A_{0}). \label{LAG5}
\end{eqnarray}
The consistency of the primary constraint with respect to the time
evolution leads to the secondary constraint
\begin{eqnarray}
\omega_{2} &=& [\pi_{0},H_{c}], \nonumber \\
&&=\pi_{1}^{\prime}+e^{2}(a-r^{2})A_{0} +e^{2}rA_{1}
+e(r\pi_{\phi}+\phi^{\prime}) \approx 0. \label{LAG7}
\end{eqnarray}
So the constraints that are embedded within the phase space of the
theory are
\begin{equation}
 \omega_{1}=\pi_{0} \approx 0, \label{LAG8}
\end{equation}
\begin{equation}
\omega_{2}=\pi_{1}^{\prime}+e^{2}(a-r^{2})A_{0}+e^{2}rA_{1}
+e(r\pi_{\phi}+\phi^{\prime})\approx 0.
 \label{LAG9}
\end{equation}
Therefore, the effective Hamiltonian in this situation reads
\begin{equation}
H_{eff}=H_{c}+u\omega_{1}+v\omega_{2}, \label{LAG10}
\end{equation}
where u and are v two lagrange multipliers. The preservation of
 $\omega_{2}$ with respect to the Hamiltonian determines the the
velocity $u$ as follows.
\begin{eqnarray}
\dot{\omega_{2}}&=&[\omega_{2},H_{eff}]\nonumber \\
&=& e^{2}(a-r^{2})u-e^{2}(a-r^{2})A_{1}^{\prime}+e^{2}r\pi_{1} =
0. \label{LAG13}
\end{eqnarray}
Equation (\ref{LAG13}) now gives
\begin{equation}
u=A_{1}^{\prime}-\frac{r}{(a-r^{2})}\pi_{1}. \label{LAG14}
\end{equation}
Therefore, substituting $u$ in (\ref{LAG10}) we get
\begin{eqnarray}
H_{eff}&=&\frac{1}{2}({\pi_{1}^{2}}+{\pi_{\phi}^{2}}
+\phi^{\prime^{2}})+ -\frac{e^{2}}{2}a(A_{0}^{2} -A_{1}^{2})
+\frac{1}{2}e^{2}(A_{1}
-rA_{0})^{2}\nonumber\\
&+& e\phi^{\prime}(rA_{1}-A_{0})
+e\pi_{\phi}(A_{1}-rA_{0})+\pi_{0}(A_{1}^{\prime}-\frac{r}{(a-r^2)}\pi_{1}).\label{LAG15}
\end{eqnarray}
The closures of the constraints with respect to the Hamiltonian
are
\begin{equation}
\dot{\omega_{1}}=[\omega_1,H_{eff}]=\omega_{2},\label{LAG16}
\end{equation}
\begin{equation}
\dot{\omega_{2}}=[\omega_{2},H_{eff}]=\omega_{1}^{\prime{\prime}}
-\frac{e^2r^2}{(a-r^2)}\omega_{1}.
\label{LAG17}
\end{equation}
For BRST invariant reformulation the system with second class
constraints (\ref{LAG16}) and (\ref{LAG17}) are needed to be
converted into a first class set. In this respect, we introduce
the auxiliary field $\theta$ and $\pi_{\theta}$. This set of
auxiliary fields satisfy the following canonical relation.
\begin{equation}
[\theta(x),\pi_{\theta}(y)]=i\delta(x-y).\label{LAG18}
\end{equation}
The auxiliary fields are known as BF fields. With some suitable
linear combinations of BF fields the second class constraints get
convert into first class constraints in the following way.
\begin{equation}
\bar{\omega_{1}}={\omega_{1}}+e(a-r^2)\theta, \label{LAG19}
\end{equation}
\begin{equation}
\bar{\omega_{2}}={\omega_{2}}+e\pi_{\theta}. \label{LAG20}
\end{equation}
For preservation of the constraints (\ref{LAG19})and
(\ref{LAG20}), $\bar{\omega_{1}}$ and $\bar{\omega_{2}}$ need to
satisfy the same closures  (\ref{LAG16}) and (\ref{LAG17}) as
satisfied by $\omega_{1}$ and $\omega_{2}$:
\begin{equation}
[H_{c},\bar\omega_1]=\bar\omega_2, \label{LAG21}
\end{equation}
\begin{equation}
[H_{c},\bar\omega_2]=\bar\omega''_1-\frac{e^2r^2}{a-r^2}\bar\omega_1.
\label{LAG22}
\end{equation}
We may expect to get first class Hamiltonian by appropriate
insertion of BF fields within the Hamiltonian (\ref{LAG15}). A
little algebra shows that the first class hamiltonian reads
\begin{equation}
\bar{H}=H_{R}+H_{BF},  \label{LAG23}
\end{equation}
where $H_{BF}$ for this theory is found out to be
\begin{equation}
H_{BF}=\frac{1}{2(a-r^2)}{\pi_{\theta}^{2}}+\frac{1}{2}(a-r^2)\theta^{\prime^{2}}
+\frac{1}{2}e^{2}r^{2}\theta^{2}.
\label{LAG24}
\end{equation}
We are now in a position to introduce the two pairs of ghost and
anti-ghost fields $(C^{i} ,\bar{P_{i}})$ and $(P^{i}
,\bar{C_{i}})$.
 We also need a pair of multiplier fields $(N_{i}, B_{i})$.
The fields satisfy the following canonical Poisson's Bracket\\
$[C^{i} ,\bar{P_{i}}]= [P^{i} ,\bar{C_{i}}]=
[N^{i},B_{j}]=i\delta^{i}_{j}\delta(x-y)$.
\label{LAG25}\\
From the definition (\ref{IM}), we can write the BRST invariant
Hamiltonian for the theory under the present situation is
\begin{equation}
H_{m}=H_{eff}+H_{BF}+\int{[Q,G]}dx+\bar{P_{a}}V^{a}_{b}C^{b}.
\label{LAG26}
\end{equation}
Here BRST charge Q and the fermionic gauge fixing function G are
defined by
\begin{equation}
Q=\int{dx(C^{1}\bar{\omega_{1}}+C^{2}\bar{\omega_{2}}+P^{1}B_{1}+P^{2}B_{2})},
\label{LAG27}
\end{equation}
\begin{equation}
G=\int{dx(\bar{C_{1}}\chi^{1}+\bar{C_{2}}\chi^{2}
+\bar{P_{1}}N^{1} +\bar{P_{2}}N^{2})}. \label{LAG28}
\end{equation}
Right now we have to  fix up the gauge condition which is very
crucial for getting appropriate Wess-Zumino term. It is found that
these two very conditions only meet our need successfully.
\begin{equation}
\chi_{1}=A_{0}, \label{LAG29}
\end{equation}
\begin{equation}
\chi_{2}=A_{1}^{\prime}+\frac{\alpha}{2} B_{2}. \label{LAG30}
\end{equation}
Let us now calculate the commutation relation in between BRST
charge, and gauge fixing function
\begin{eqnarray}
[Q,G]&=&[B_{i}P^{i}+C^{i}\bar{w_{i}},\bar{C_{j}}\chi^{j}+\bar{P_{j}}N^{j}]\nonumber\\
&=&B_{1}\chi^{1}+B_{2}\chi^{2}
-P^{1}\bar{P_{1}}-P^{2}\bar{P_{2}}-C^{1}\bar{C_{1}}
+C^{2}\bar{C_{2}^{\prime\prime}}\nonumber\\
&+&\bar{\omega_{1}}N^{1}+\bar{\omega_{2}}N^{2}. \label{LAG31}
\end{eqnarray}
Using equation (\ref{LAG26}), BRST invariant Hamiltonian is
obtained  which is given by
\begin{equation}
H_{m}=H_{R}+H_{BF}+\bar{P_{2}}C_{1}
+\bar{P_{1}}^{\prime\prime}C_{2}-\frac{e^{2}r^{2}}{a-r^{2}}\bar{P_{1}}C_{2}+\int[Q,G]dx.\label{LAG32}
\end{equation}
The generating functional for this system can now be written down
as
\begin{equation}
Z=\int[D\mu]exp^{iS}, \label{LAG33}
\end{equation}
where $[D\mu]$ is the Liouville measure in the extended phase
space.
\begin{equation}
[D\mu]=d\phi d\pi_{\phi} dA_{0} d\pi_{0} dA_{1} d\pi_{1} d\theta
d\pi_{\theta} \prod_{i=1}^2 dN^{i} dB_{i}  dC^{i}, d\bar C_{i}
dP^{i}d\bar P_{i}. \label{LAG34}
\end{equation}
The action $S$ in equation (\ref{LAG33}) reads
\begin{eqnarray}
S&=& \int d^{2}x [\pi_{\phi}\dot{\phi}+\pi_{0}\dot{A_{0}}
+\pi_{1}\dot{A_{1}} +\dot{\theta}\pi_{\theta}
+\dot{N^{1}}B_{1}\nonumber\\
&+&\dot{N^{2}}B_{2}+ \bar C_1\dot P^1+\bar C_2\dot P^2 + \bar P^1
\dot C_1 +\bar P^2 \dot C_2  - H_{m}]. \label{LAG35}
\end{eqnarray}
The explicit form of  $H_{m}$ lying in  equation (\ref{LAG35}) is
\begin{eqnarray}
H_{m}&=&
H_{R}+H_{BF}-P^{1}\bar{P_{1}}-P^{2}\bar{P_{2}}-C^{1}\bar{C_{1}}
+C^{2}\bar{C_{2}^{\prime\prime}}\nonumber\\
&+&\bar{\Omega_{1}}N^{1}+\bar{\Omega_{2}}N^{2} +\bar{P_{2}}C_{1}
+\bar{P_{1}}^{\prime\prime}C_{2}\nonumber\\
&-&\frac{e^{2}r^{2}}{a-r^{2}}\bar{P_{1}}C_{2} +B_{1}A^{0}
+B_{2}(A_{1}^{\prime} +\frac{\alpha}{2}B_{2}). \label{LAG36}
\end{eqnarray}
To get the effective action in the desired form it is necessary to
integrate out the fields
 $B_{1},N^{1},\pi_{0}, \pi_{1},\pi_{\phi}, \bar{P_{1}}, C_{1}$ and $\bar{C_{1}}$.
After integrating out of these fields we obtain a simplified form
of the generating functional which is constituted with the
following action:
\begin{eqnarray}
S&=&\frac{1}{2}\partial_{\mu}\phi\partial^{\mu}\phi
+e\epsilon_{\mu\nu}A^{\mu}\partial^{\nu}\phi
+erg_{\mu\nu}A^{\mu}\partial^{\nu}\phi +\frac{1}{2}ae^2A_{\mu}A^{\mu}
-\frac{1}{4}F_{\mu\nu}F^{\mu\nu} \nonumber\\
&+&\frac{1}{2}(a-r^{2})\partial_{\mu}\theta\partial^{\mu}\theta
+e(a-r^{2})g_{\mu\nu}A^{\mu}\partial^{\nu}\theta -
er\epsilon_{\mu\nu}A^{\mu}\partial^{\nu}\theta \nonumber\\
&+& \partial^{\mu}C\partial_{\mu}\bar{C}
+\frac{\alpha}{2}B^2+B\partial_{\mu}A^{\mu}. \label{LAG37}
\end{eqnarray}
We have to choose  $ C^{2}=C, N^{2}=A_{0}, B_{2}=B $ to reach
equation (\ref{LAG37}) from equation (\ref{LAG33}). It is
interesting to see that the action (\ref{LAG37}) is  invariant
under the transformation
\begin{equation}
\delta{\phi}=er\lambda C,\delta N_{0}=-\lambda\dot{C},
\delta{A_{1}}=-\lambda C^{\prime}, \delta{\theta}=-\lambda eC,
\delta C=0, \delta \bar{C}=-\lambda B. \label{LAG38}
\end{equation}
These are the very BRST transformations corresponding to the
fields that describe the  system under consideration. It would be
of worth to reiterate that the choice of gauge fixing is very
crucial here. The choice of gauge fixing which we have considered
here renders a great service to obtain the appropriate Wess-Zumino
term. The Wess-Zumino term $L_{wz}$ can easily be identified as
\begin{equation}
L_{wz}=\frac{1}{2}(a-r^{2})\partial_{\mu}\theta\partial^{\mu}\theta
+e(a-r^{2})g_{\mu\nu}A^{\mu}\partial^{\nu}\theta
-er\epsilon_{\mu\nu}A^{\mu}\partial^{\nu}\theta.
\label{LAG39}
\end{equation}
This completes the BRST invariant reformulation of the theory. we
will now proceed to the application of FFBRST and anti-FFBRST
formalism on the same model in the next section.
\section{Application of FFBRST and anti-FFBRST formalism in the GVSM}
An ingenious attempt was made in \cite{JOG} to generalize the well
celebrated BRST formulation. It was shown there that even making
the BRST transformation field dependent the nilpotency can be
protected and it is equally effective for anti BRST formalism.
Under finite field dependent transformation the path integral
measure acquires a nontrivial change that though leads to a
different effective theory, the physical contents of the theory
remains unaffected. This generalization however is advantageous
since that renders several important services. One of such
advantage is that it helps to correlate the different gauge fixed
versions of a particular theory \cite{JOG}. The ability to relate
a theory embedded with a set of first class constraint to an
equivalent theory embedded with a set of second class constraint
through appropriate choice of gauge fixing parameter is also an
interesting extension of the field dependent BRST (FFBRST)
formalism.  \cite{BHS, BHS1, BHS2, BHS3, BHS4,
BHS5,BHS6,BHS7,BHS8,BHS9,BHS10}.

 An illustration related to the calculation of jacobian for this
 field dependent transformation is available in \cite{JOG}. To make this paper
a self contained one let us now
 proceed  with the brief
introduction concerning how FFBRST transformation brings a non
trivial change in the interactional measure of the generating
functional and how this change adds a contribution to the
effective action.

If the fields that describe a physically sensible theory are
function of a
 parameter $\eta$ such that $\phi(x, \eta)$ is defined by $\phi(x,
 \eta=0)=\phi(x)$ and $\phi(x, \eta=1) =\tilde\phi(x)$ the
 infinitesimal BRST transformation is given by
\begin{equation}
\frac{d}{d\eta}\phi(x) =
\delta_{BRS}[\phi(\eta)\Theta'[\phi(\eta)].
 \end{equation}
 The finite field dependence can be obtained through the
 integration over the infinitesimal transformation within the
 limit $\eta=0$ to $\eta=1$
\begin{equation}
\tilde\phi(x)=\phi(x, \eta=1)=\phi(x, \eta=0)+
\delta_{BRS}[\phi(x)\Theta[\phi(x)],
\end{equation}
where
\begin{equation}
 \Theta[\phi(x)] =\int_0^1 d\eta'\Theta'[\phi(x, \eta')].
\end{equation}
It should be mentioned here that the condition  $\Theta^2=0$ is to
be maintained in order to protect nilpotency. The jacobian for the
transformation can be evaluated from the field dependent function
$\Theta[\phi(x)]$ by
\begin{equation}
\prod d\phi = J(\eta)\prod d\phi(\eta)= J(\eta+ d\eta)\prod
d\phi(\eta+ d\eta).
\end{equation}
The infinitesimal nature of transformation from $\phi(\eta)$ to
$\phi(\eta +d\eta)$ leads to the following relations of the
jacobian $J(\eta)$:
\begin{equation}
 \frac{J(\eta)}{J(\eta+d\eta)} = \sum \pm\frac{\delta\phi(x, \eta
 +d\eta)}{\delta\phi}. \label{JC}
\end{equation}
Here $\prod$ and $\sum$ represents the product and sum over all
the fields involved within the theory respectively. In equation
(\ref{JC}) $(+)$ and $(-)$ signs are to be used for boson and
fermion field respectively. Equation (\ref{JC}) renders the
following in infinitesimal change in the jacobian $J(\eta)$.
\begin{equation}
 \frac{1}{J} \frac{dJ}{d\eta} = -\int d^2 x[\pm \delta_b\phi(x,
 \eta)\frac{\partial\Theta'}{\partial \phi}]\label{JC1}.
\end{equation}
The incredible characteristic of this extension is that within the
functional integration $J(\eta)$ can be expressed as
\begin{equation}
 J(\eta) = e^{iS_c(x, \eta)},
\end{equation}
 {\it if and only if} the condition
\begin{equation}
 \int \prod d\phi(x)[\frac{1}{J} \frac {dJ}{d \eta} - i \frac{dS_C(x,
 \eta)}{d\eta}]e^{i(S_eff + S_C)} =0,
\end{equation}
 is maintained within the phase space of the theory.
The role of $\Theta$ though surprising, nevertheless plays a very
crucial as well as intriguing role since  the appropriate choice
$\Theta$ leads to another equivalent effective action
corresponding to the starting theory which is  given by the
geherating functional.
\begin{equation}
\tilde Z  = \int \prod d\phi(x)\exp i(S_{eff} + S_C).
\end{equation}
It indeed keeps the physical contents of the theory unchanged.

To see the transmutation between the gauge invariant and gauge
non-invariant effective theory of GVSM we begin our analysis
starting from the BRST invariant effective action of the theory
which reads
\begin{eqnarray}
S_{eff}&=&\frac{1}{2}\partial_{\mu}\phi\partial^{\mu}\phi
+e\epsilon_{\mu\nu}A^{\mu}\partial^{\nu}\phi
+erg_{\mu\nu}A^{\mu}\partial^{\nu}\phi
+\frac{1}{2}ae^2A_{\mu}A^{\mu}
-\frac{1}{4}F_{\mu\nu}F^{\mu\nu} \nonumber\\
&+&\frac{1}{2}(a-r^{2})\partial_{\mu}\theta\partial^{\mu}\theta
+e(a-r^{2})g_{\mu\nu}A^{\mu}\partial^{\nu}\theta -
er\epsilon_{\mu\nu}A^{\mu}\partial^{\nu}\theta \nonumber\\
&+& \partial^{\mu}C\partial_{\mu}\bar{C}
+\frac{\alpha}{2}B^2+B\partial_{\mu}A^{\mu}
\nonumber \\
&=&S_{ORI} + S_{WZ} +  S_{GHOST} + S_{GF}. \label{SEFF}
\end{eqnarray}
The infinitesimal BRST transformations of the fields under which
the above action (\ref{SEFF}) is found to  remain invariant are
\begin{equation}
\delta A_\mu = -\frac{1}{e} \lambda \partial_\mu C, \delta
\phi=r\lambda C, \delta \theta= \lambda C, \delta C =0, \delta
\bar C =-B,\delta B =0,
\end{equation}
and the FFBRST transformations  of those fields describing the
theory under consideration  are
\begin{equation}
\delta A_\mu = -\frac{1}{e}  \partial_\mu C \Theta, \delta \phi=r
C\Theta, \delta \theta= -\Theta C, \delta C =0, \delta \bar C
=-B\Theta,\delta B =0,
\end{equation}
where $\Theta$ is an arbitrary finite field dependent function
serving the role of transformation parameter corresponding to the
FFBRST transformation. Our objective is to connect this BRST
invariant effective action of the extended phase space to the
original effective action of the usual physical phase space. To
relate this we make a choice over the $\Theta$ in such a way that
the change that would enter into the measure of the generating
functional can be exploited to serve the desired purpose. To this
end we define $\Theta$ as follows:
\begin{equation}
\Theta'= i\gamma\int d^2 x[\bar C(\partial_\mu A^\mu +
\frac{\alpha}{2} B)]. \label{DTH}
\end{equation}
Here $\gamma$ is an arbitrary parameter that would be fixed later.
For the finite field dependent parameter the nontrivial
infinitesimal change that would enter in the jacobian can be be
computed using equation (\ref{JC1})
\begin{equation}
\frac{1}{J}\frac{dJ}{d\eta} =i\gamma\int d^2 x[\delta\bar
C\frac{d\Theta'}{d\bar C}+\delta A_\mu\frac{d\Theta'}{d A_\mu} +
\delta B\frac{d\Theta'}{dB}]
\end{equation}
\begin{equation}
= i\gamma\int d^2 x [B(\partial_\mu A^\mu + \frac{\alpha}{2}
B)+\frac{\lambda}{e}\partial_\mu C\partial^\mu \bar C].
\end{equation}
The Euler-Lagrange  equation of motion for the ghost field
simplifies the above equation to the following form
\begin{equation}
\frac{1}{J}\frac{dJ}{d\eta} = i\gamma\int d^2 x[ -B(\partial_\mu
A^\mu + \frac{\alpha}{2} B)].
\end{equation}
We are now in a state to choose an ansatz for $S_C$. The following
ansatz for $S_C$ suffices our need without violating any physical
principle
\begin{equation}
S_C = \int d^2 x [\zeta_1(k)B^2 +\zeta_2(k)B\partial_\mu A^\mu].
\end{equation}
Here $\zeta_1(\eta)$ and $\zeta_2(\eta)$ are some function of the
parameter $\eta$. The differentiation  of the action $S_C$ with
respect to $\eta$ yields
\begin{equation}
 \frac{\partial S_C}{\partial \eta} = \int d^2 x( B^2\zeta'_1(\eta) + B\partial_\mu
 A^\mu\zeta'_2(\eta)).
\end{equation}
The over prime denotes here the differentiation with respect to
the parameter $\eta$. The contribution that enters into the
measure of the generating function through the jacobian under
FFBRST transformation can be written down in the form of
$e^{iS_C}$, provided the following very equation
\begin{equation}
\int d^2 x e^{i(S_{eff} + S_C)}[ i(\zeta'_1 - \frac{\gamma
\alpha}{2}B^2)+ iB(\partial_\mu A^\mu( \zeta'_2 - \gamma)] =0,
\end{equation}
is satisfied. It fixes  $\zeta_1$ and $\zeta_2$ and make it
expressible in terms of $\gamma$ and $\alpha$:
\begin{eqnarray}
\zeta_1&=& \frac{\gamma \alpha}{2}\eta \nonumber \\
\zeta_2 &=& \gamma \eta.
\end{eqnarray}
Setting $\eta=1$, we get
\begin{equation}
S_C = \int d^2 x [\frac{\gamma \alpha}{2} B^2 +\gamma
B\partial_\mu A^\mu],
\end{equation}
and for $\gamma$ =-1,  $S_C$ turns into
\begin{equation}
 S_C = \int d^2
x [-(\frac{\alpha}{2} B^2 +B\partial_\mu A^\mu)].
\end{equation}
So through the exploitation of the change entered in to the
measure of the generating functional through FFBRST transformation
enables us to eliminate the gauge fixing part $S_{GF}$ from the
$S_{eff}$ for the above setting of $\gamma$ and $\eta$.  It is the
first step to proceed towards the effective action defined in the
usual phase space. So the remaining part in the $S_{eff}$ are
\begin{equation}
S_{ST} = S_{ORI} + S_{WZ} +  S_{GHOST}.
\end{equation}
Precisely, the parameter of FFBRST transformation (being field
dependent) renders here a great job which is the elimination of
the gauge fixing term through the contribution entered into the
path integral measure of generating functional due to the finite
field dependent nature of the transformation parameter of the
FFBRST transformation. Our next task is to eliminate the ghost and
the Wess-Zumino part one by one. The elimination of the  ghost
part is trivial because under integration the contribution that
evolve out from this part can be  absorbed within the
normalization. However, the  elimination of the Wess-Zumino term
is not so trivial. It certainly needs the integrating out of the
Wess-Zumino field but one has to keep it in mind that the theory
has now converted into  gauge invariant one and the constrains
that embedded in its phase space are first class in nature. So
proper gauge fixing is needed to land onto the theory of the usual
phase pace \cite{FALCK}. This can be done in different ways.  In
\cite{FALCK}, the authors did not use the path integral approach.
However, since in the full body of the paper path integral
approach is followed we use gauge fixing with the path integral
formulation to which we now tern.

From the Hamiltonian analysis which is  available in \cite{ARSN},
it is known that the phase space of the theory contains two first
class constraint. From \cite{ARSN}, we find that the the original
action along with the Wess-Zumino part of the
$S_{eff}$(\ref{SEFF}), leads to the following Hamiltonian.
\begin{eqnarray}
H_{ce}&=& \int dx[\frac{1}{2}(\pi_\phi^{2}+\pi_1^2
+\phi'^2)+\pi_1A'_0+\frac{e^{2}a(1+a-r^{2})}{2(a-r^{2})}A_1^2
+e\phi'(rA_1-A_0)\nonumber\\
&+& e\pi_\phi(A_1-rA_0) +\frac{1}{2}(a-r^2)\theta'^2
+e\theta'((a-r^2)A_{1}+A_{0}) +\frac{1}{2(a-r^2)}{\pi_{\theta}^{2}}\nonumber\\
&+&\frac{e}{(a-r^{2})}(rA_{1}+(a-r^{2})A_{0})\pi_{\theta}],
\label{CHAM}
\end{eqnarray}
and there embeds the following two first class constraints in the
phase space of the theory:
\begin{equation}
\tilde\omega_{1}=\pi_{0}\approx 0, \label{CON1}
\end{equation}
\begin{equation}
\tilde\omega_{2}= \pi_{1}^{\prime}+e(r\pi_\phi+
\phi')+e(\pi_\theta- r\theta')\approx 0 . \label{CON2}
\end{equation}
Therefore, two gauge fixing conditions are needed at this stage to
get back the gauge non-invariant theory of the usual phase space.
These two gauge fixing conditions that are chosen here are
\begin{equation}
\tilde\omega_{3}=\theta'\approx 0, \label{GF1}
\end{equation}
\begin{equation}
\tilde\omega_{4}=\pi_{\theta}+e((a-r^{2})A_{0}+rA_{1})\approx 0.
\label{GF2}
\end{equation}
With these inputs the generating functional now can  be written
down as
\begin{equation}
Z=
\int[D\mu][det[\tilde\omega_i,\tilde\omega_j]^{\frac{1}{2}}e^{i\int
d^2x[\pi_1\dot A_1 + \pi_0 \dot A_0 +\pi_\phi \dot\phi+ \pi_\theta
\dot\theta - H_{ce}]} \delta(\tilde\omega_1)(\tilde\omega_1)
\delta(\tilde\omega_3)\delta(\tilde\omega_4)\label{CZ},
\end{equation}
where the Liouville measure $[D\mu] = d\pi_\phi d\phi d\pi_1dA_1
d\pi_0 dA_0 d\pi_\theta d\theta$, and $i$ and$j$ runs from $1$ to
$4$. After integrating out of the field $\theta$ and
$\pi_{\theta}$ we find that equation (\ref{CZ}) turns into
\begin{equation}
Z= N \int[d\bar\mu]e^{i\int d^2x[\pi_1\dot A_1 + \pi_0 \dot A_0
+\pi_\phi \dot\phi+  - \bar H_{ce}]} \delta(\pi_0) \delta(\pi'_1+
e(r\pi_{\phi}+ \phi') +e^2((a-r^{2})A_{0}+rA_{1}),\label{CCZ}
\end{equation}
where $[d\bar\mu] = d\pi_\phi d\phi d\pi_1dA_1 d\pi_0 dA_0 $ and
$N$ is a normalization constant having no significant physical
importance and $\tilde H_{ce}$ is
\begin{eqnarray}
\tilde H_{ce}=\frac{1}{2}(\pi_1^2+\phi'^2)+\pi_1A_0'
+\frac{1}{2}ae^2 (A_1^2-A_0)^2
+e\phi'(rA_1-A_0)+\frac{1}{2}[\pi_\phi+e(A_1-rA_0)]^2.\label{BARH}
\end{eqnarray}
Now after integrating out of the the field $\pi_\phi$, $\pi_1$ and
$\pi_0$ we land on to the required result
\begin{equation}
Z=\int dA_1 dA_0 d\phi e^{iS_{ORI}}.
\end{equation}

Like BRST symmetry anti-BRST is also a symmetry of the effective
action of a given theory and like the BRST transformations
anti-BRST transformations do generate from a nilpotent charge. In
the  anti-BRST formulation the role of ghost and anti ghost fields
interchanges. In addition to that there may be change in the
coefficient depending upon the system. Therefore, study with
anti-FFBRST is equally important like FFBRST. So we are intended
to examine whether the anti-FFBRST formalism can be brought into
the same service as it has been served by the FFBRST formalism.
The ant-BRST transformations for the fields describing the theory
are given by
\begin{equation}
\delta A_\mu = -\frac{1}{e} \lambda \partial_\mu \bar C, \delta
\phi=r\lambda \bar C, \delta \theta= -\lambda \bar C, \delta \bar
C =B\lambda \delta \bar C =0,\delta B =0,
\end{equation}
and the corresponding  anti-FFBRST transformations of the fields
are
\begin{equation}
\delta A_\mu = -\frac{1}{e}  \partial_\mu \bar C \Theta_A, \delta
\phi=r\bar C\Theta_A, \delta \theta= - \bar C \Theta_A, \delta
\bar C =0, \delta C =-B\Theta_A,\delta B =0,
\end{equation}
where $\Theta_A$ is an arbitrary finite field dependent function
serving the role of transformation parameter corresponding to the
anti-FFBRST transformation. Our objective is the same as we have
done for the FFBRST. For that purpose here also we need to choose
a $\Theta_A$ in such a way that the change that would enter into
the generation functional can be exploited to serve the same
purpose as it has been found to serve in the earlier situation. In
an analogous manner let us define $\Theta_A$ as
\begin{equation}
\Theta'_A= i\tilde\gamma\int d^2 x[ C(\partial_\mu A^\mu +
\frac{\beta}{2} B)], \label{ADTH}
\end{equation}
here $\tilde\gamma$ is an arbitrary parameter that would be fixed
later like the previous situation. For anti-FFBRST transformation
also the nontrivial infinitesimal change in the jacobian that
would enter can be calculated using equation (\ref{JC1}).
\begin{equation}
 \frac{1}{J} \frac{dJ}{d\eta} = i\gamma\int d^2 x [B(\partial_\mu A^\mu + \frac{\beta}{2} B)
+\frac{\lambda}{e}\partial_\mu C\partial^\mu \bar C].
\end{equation}
By the use of Euler-lagrange  equation of motion for the anti
ghost field the above equation reduces to
\begin{equation}
 \frac{1}{J} \frac{dJ}{d\eta} = i\tilde\gamma\int d^2 x[ -B(\partial_\mu A^\mu + \frac{\beta}{2}
B)].
\end{equation}
In order to express the above in the form of $e^{i\tilde S_C}$,
the following ansatz can be chosen for $\tilde S_C$  without any
loss of generic condition, and of course, without violating any
physical principle:
\begin{equation}
\tilde S_C = \int d^2 x [\xi_1(k) B^2 +\xi_2(k)B\partial_\mu
A^\mu].
\end{equation}
Here $\xi_1(\eta)$ and $\xi_2(\eta)$ are some function of the
parameter $\eta$. If we now take the derivative of  the action
$\tilde S_C$ with respect to $\eta$ we get
\begin{equation}
 \frac{\partial S_C}{\partial \eta} =\int d^2 x [B^2\xi'_1(\eta)
 + B\partial_\mu A^\mu\xi'_2(\eta)].
\end{equation}
The symbol over prime denotes here the differentiation with
respect to the parameter $\eta$ as usual. The contribution that
the path integral measure of the generating functional acquires
under anti-FFBRST transformation can be written down in the form
of $e^{i\tilde S_C}$, {\it iff} the following relation
\begin{equation}
\int d^2 x e^{i(S_{eff} + S_C)}[ i(\xi'_1 - \frac{\tilde\gamma
\beta}{2}B^2)+ iB(\partial_\mu A^\mu( \xi'_2 - \tilde \gamma)]
=0,\label{IFF}
\end{equation}
holds.  The equation (\ref{IFF}) fixes $\xi_1$ and $\xi_2$ so one
can express these in terms of $\tilde \gamma$ and $\beta$:
\begin{eqnarray}
\xi_1&=& \frac{\tilde\gamma \beta}{2}\eta, \nonumber \\
\xi_2 &=& \tilde\gamma \eta.
\end{eqnarray}
Thus setting the parameter $\eta=1$ we get
\begin{equation}
\tilde S_C = \int d^2 x [\frac{\tilde\gamma \beta}{2} B^2
+\tilde\gamma B\partial_\mu A^\mu],
\end{equation}
and finally putting $\tilde\gamma$ =-1, we get the appropriate
$\tilde S_C$ in his situation:
\begin{equation}
 \tilde S_C =
 \int d^2 x [-(\frac{\beta}{2} B^2 +B\partial_\mu A^\mu)].
\end{equation}
Therefore, we find that the exploitation of the change entered
into path integral measure of the generating functional due to the
anti-FFBRST transformation with the above choice of $\tilde\gamma$
and $\eta$ enables us to eliminate the gauge fixing term from our
starting $S_{eff}$ and it now reduces to
 \begin{equation} S_{eff} = S_{ORI} + S_{WZ}
+  S_{GHOST}.
\end{equation}
So the first step to reach towards the effective action defined n
the usual phase space is successfully made in case of anti-FFBRST
transformation too. Note that for this system the calculation may
look similar to the FFBRT. However, for the other situations this
task may not be so straightforward. For instance, the BRST and
ant-BRST transformation for the gauge fixed Yang-Mills theory
differs in coefficient \cite{JOG}, and as a result the
calculations in this regard takes a different shape.

 After the first step, the task that is yet to be done is to
 make the $S_{eff}$ part free from ghost as well as the Wess-Zumino part.
 The elimination of the ghost part is trivial like the previous case
 since under integration of the anti-ghost field the contribution that evolve
  out can again be absorbed within the normalization. However, the
  elimination of the Wess-Zumino term is not trivial but it is
  identical to the previous case as it has already been made for
  the FFBRST transformation. Explicit calculation in this situation
  therefore does not carry any new information. So, it is not shown here.

\section{conclusion}
The present work is an application of the well celebrated BRST
formulation and its incredible generalization, so called FFBRST
and anti-FFBRST formulation on a lower dimensional model where
vector and axial vector coupling gets mixed up with different
weight \cite{BAS1, BAS2}. Initially, we have attempted to
construct the BRST invariant reformulation using the improved
version of BFV formulation due to Fujiwara, Igarishi and Kubo
\cite{FIK}. This improved version has helped us to obtain a BRST
invariant reformulation along with the emergence of appropriate
Wess-Zumino term. We would like to emphasize the fact that the
role of gage fixing is very crucial to get the appropriate
Wess-Zumino term.

Application of FFBRST and anti-FFFBRST are made for another
interesting as well as important purpose. In fact, it has been
used here to make the equivalence between the physical content of
the model in the usual and in the extended phase space. Here
extended phase space implies the presence of not only the
Wess-Zumino field, but also the presence of ghost and the
auxiliary $B$ fields too. It has been found that both the FFBRST
and anti-FFBTST formulation have successfully render their great
services to show the equivalence.

In both the cases FFBRST and ant-FFBRST it has been found that the
gauge fixing part, i.e., the part of the effective action
involving the auxiliary $B$ gets eliminated by the contribution
entered into the effective action through the acquired
contribution of the measure of the generating functional under
FFBRST and anti-FFBRST transformations respectively. To eliminate
rest of the part we have adopted here the formalism developed by
Falck and Kramer \cite{FALCK}.  So, the joint action of the two
formalisms developed in \cite{JOG} and \cite{FALCK} have done
their novel services to show the equivalence. We can conclude that
the joint action of these two formalisms would be instrumental to
show the equivalence between the different effective actions of
any field theoretical model if these two are employed in
appropriate manner.

\end{document}